\documentclass[aps,prl,twocolumn,showpacs]{revtex4}
\usepackage{stmaryrd}
\usepackage{amssymb}

\usepackage{pifont}
\usepackage{amsfonts}
\usepackage{graphicx}

\begin{document}

\title{Synthesizing and characterization of hole doped nickel based superconductor (La$_{1-x}$Sr$_{x}$)NiAsO}

\author{Lei Fang, Huan Yang, Peng Cheng, Xiyu Zhu, Gang Mu }
\author{Hai-Hu Wen}\email{hhwen@aphy.iphy.ac.cn}

\affiliation{National Laboratory for Superconductivity, Institute of
Physics and Beijing National Laboratory for Condensed Matter
Physics, Chinese Academy of Sciences, P.O. Box 603, Beijing 100190,
People's Republic of China}

\begin{abstract}
We report the synthesizing and characterization of the hole doped
Ni-based superconductor ($La_{1-x}Sr_{x})NiAsO$. By substituting La
with Sr, the superconducting transition temperature $T_c$ is
increased from 2.4 K of the parent phase $LaNiAsO$ to 3.7 K at the
doping levels x= 0.1 - 0.2. The curve $T_c$ versus hole
concentration shows a symmetric behavior as the electron doped
samples $LaNiAs(O_{1-x}F_{x})$. The normal state resistivity in
Ni-based samples shows a good metallic behavior and reveals the
absence of spin density wave induced anomaly which appears in the
Fe-based system at about 150 K. Hall effect measurements indicate
that the electron conduction in the parent phase $LaNiAsO$ is
dominated by electron-like charge carriers, while with more Sr
doping, a hole-like band will emerge and finally prevail over the
conduction, such a phenomenon reflects that the Fermi surface of
$LaNiAsO$ comprises of electron pockets and hole pockets, thus the
sign of charge carriers could be changed once the contribution of
hole pockets overwhelms that of electron pockets. Magnetoresistance
measurements and the violation of Kohler rule provide further proof
that multiband effect dominate the normal state transport of
($La_{1-x}Sr_{x})NiAsO$.

\end{abstract}
\pacs{74.10.+v, 74.70.Dd, 74.20.Mn}

\maketitle

 Searching for high temperature superconductors has been a
 long-term strategy in material science. Superconductors
 with unconventional pairing symmetry
 found in past decades seem to
 have some common features: layered structure, such as in
 cuprates\cite{1}; tunable transition
 temperature (T$_{c}$) by doping holes or electrons; possible exotic pairing mechanism rather than phonon mediated
 superconductivity, for instance in the heavy fermion system\cite{2}.  The newly discovered
 iron-based
 superconductor LaFeAs(O$_{1-x}$F$_{x}$) with a moderate high T$_c$ = 26 K
 seems to fit to these three categories\cite{3,4,MuG,ShanL}. It is found that LaFeAs(O$_{1-x}$F$_{x}$)
 belongs to a layered structure constructed by stacking the LaO and FeAs sheets alternatively, where FeAs sheet is regarded as the conduction layer
 whose charge carrier density could be tuned by the neighboring LaO
 sheet by charge doping. Substituting part of the oxygen with fluorine, the system
 changes from having a weak insulating behavior to superconductive with x = 0.05 - 0.12\cite{4}. This discovery
 has stimulated intense efforts in both experimental and theoretical
studies.  Theoretically it was concluded that the electronic
correlation of this system could be moderate\cite{5,6}.
Experimentally both low temperature specific heat\cite{MuG} and
point contact tunneling\cite{ShanL} measurements indicate the
possible unconventional pairing symmetry. LnFeAsO$_{1-x}$F$_{x}$ (Ln
represents the rare earth element La, Ce, Pr, Nd, Sm, Gd) has been
proved to be bearing electron type carriers, thus the possibility to
realize hole doped superconductor in such a system is very
attractive. The first stride has been successfully made in hole
doped samples ($La_{1-x}Sr_x)FeAsO$ with T$_{c}$ = 25K by our
group\cite{7}. More recently hole doped La$_{1-x}$Ca$_{x}$FePO and
(Ba$_{1-x}$K$_{x}$)Fe$_{2}$As$_{2}$ have been reported to realize
 superconductivity\cite{hosono ca, Johrendt}. In present
work we report the fabrication and characterization of the hole
doped Ni-based superconductors (La$_{1-x}$Sr$_{x}$)NiAsO.
Superconductivity at about 3.7 K was found and the $T_c$ exhibits a
symmetric behavior in both hole-doped and electron-doped side. The
hole-like charge
     carriers in the present Sr doped sample (high doping) are evidenced by Hall effect measurements.

\begin{figure}

       \includegraphics[width=8.2cm]{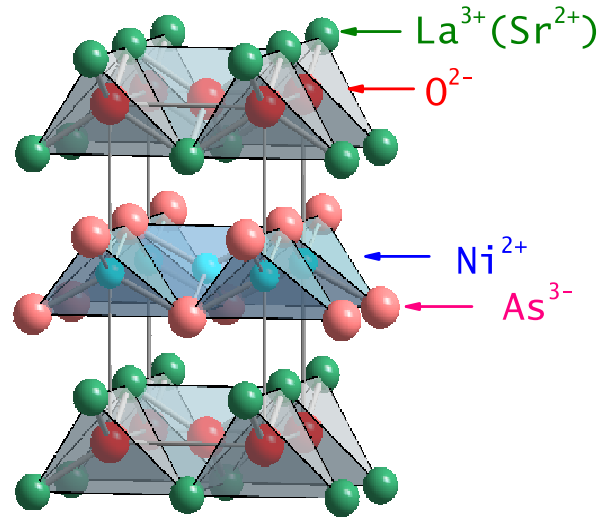}
       \caption{(Color online) Schematic illustration of (La$_{1-x}$Sr$_{x}$)NiAsO
       structure. Parts of La$^{3+}$ coordinations were occupied by
       Sr$^{2+}$, thus hole was implanted into parent phase LaNiAsO.
     }
       \label{figure1}
\end{figure}

Polycrystalline  (La$_{1-x}$Sr$_{x}$)NiAsO samples ( x = 0.1, 0.2,
0.3) were synthesized by the conventional solid state reaction
method. Stoichiometric LaAs powder was home made by reacting pure
$La$ (99.99\%) and $As$ (99.99\%). Later it is mixed with dehydrated
La$_{2}$O$_{3}$(99.9\%), SrO (99.5\%), and NiAs powder (home made by
reacting pure $Ni$ (99.99\%) and $As$ (99.99\%)),
     and Ni powder (99.99\%), grounded and pressed into a
     pellet. Then the pellet was sealed into an evacuated quartz tube.
     Consequently, the tube was slowly warmed up in a muffle furnace to 1150 $^{\circ}$ C
     and sintered for 48 hours, then cooled down to room
     temperature. X-ray diffraction (XRD) pattern measurement was performed at room temperature employing an M18AHF
     x-ray diffractometer (MAC Science).
     The magnetic measurements
     were carried out on a Magnetic Property Measurement System (MPMS, Quantum Design). The electrical resistivity and Hall coefficient
     were measured by a six-probe method based on a Physical Property Measurement System (PPMS, Quantum Design).

\begin{figure}

       \includegraphics[width=9cm]{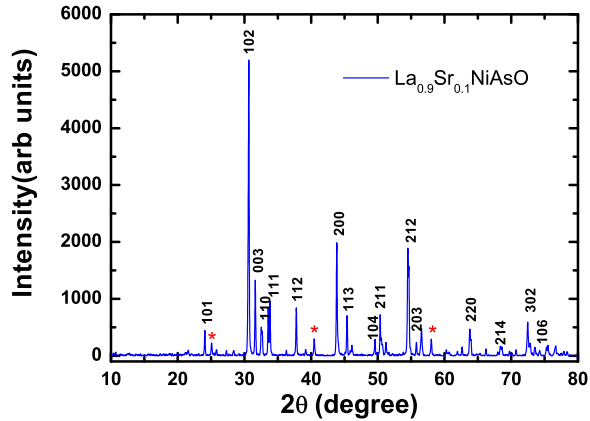}
       \caption{(Color online) XRD pattern of (La$_{1-x}$Sr$_{x}$)NiAsO with x = 0.1, which can be indexed in a
     tetragonal symmetry with a = b = 4.1290 ${\AA}$ and c = 8.1936 ${\AA}$. The asterisks mark the peaks from impurity phase.
     }
       \label{figure2}
\end{figure}

     Fig.1 presents the schematic illustration of (La$_{1-x}$Sr$_{x}$)NiAsO
       structure. Parts of La$^{3+}$ coordinations were occupied by
       Sr$^{2+}$. Fig.2  shows the XRD pattern of the
     sample (La$_{0.9}$Sr$_{0.1}$)NiAsO, which can be indexed in a
     tetragonal space group with \emph{a} = \emph{b} = 4.1290 ${\AA}$ and \emph{c} = 8.1936 ${\AA}$.
     Though minor peaks arising from the impurity phase were found (could come from NiAs),
     there is no doubt that the main phase is dominated by (La$_{1-x}$Sr$_{x}$)NiAsO in the sample with x = 0.10.
     Crystalline quality of (La$_{0.8}$Sr$_{0.2}$)NiAsO is similar to that
     of 0.1 doping but with a bigger lattice parameter(\emph{a} = \emph{b} = 4.1483 ${\AA}$ and \emph{c} =
     8.2105${\AA}$). Comparing to the indices of parent phase
     LaNiAsO(\emph{a} = \emph{b} = 4.12309${\AA}$ and \emph{c} = 8.18848 ${\AA}$)\cite{hosono LaONiAs}, cell parameters of strontium doped (La$_{1-x}$Sr$_{x}$)NiAsO
     are a bit larger. As to the 0.3 doping, it is observed that
     lots of impurity peaks  dominated the XRD pattern, indicating stronger phase segregation during the
     sintering. Taking account of the sandwich structure of LaNiAsO,
     structure distortion of LaO sheet caused by incommensurate replacement of La$^{3+}$ by
     Sr$^{2+}$  is restricted by the neighboring NiAs sheets, thus
     it is believed that the quantity of chemical doping is limited
     on certain extent. Therefore in the following discussion,
     the data of (La$_{0.7}$Sr$_{0.3}$)NiAsO are not included. An
      interesting result is that the cell parameters increase with
      the Sr doping, meanwhile T$_{c}$ also increases and saturates to
      high doping(shown in the following section). However, electron
      doped (fluorine) and hole doped (calcium) materials have shown
      that T$_{c}$ is proportional to the shrinkage of cell parameters\cite{hosono ca}.
      It should be pointed that La$^{3+}$ and Sr$^{2+}$
      ions have radius of 1.06${\AA}$ and 1.12${\AA}$,
      respectively. The size difference is not that big. Thus we suggest that for hole doped
      LaNiAsO
      chemical pressure could not be the only parameter to
      influence T$_{c}$, the band filling may also play an important role.

\begin{figure}

       \includegraphics[width=8cm]{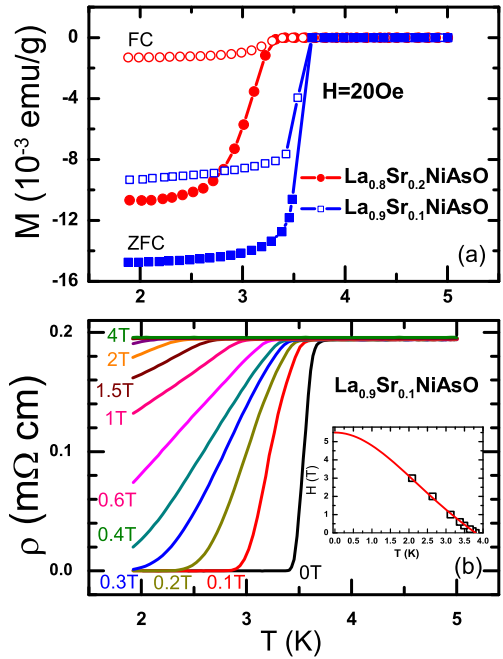}
       \caption{(Color online) (a) DC magnetization of (La$_{1-x}$Sr$_{x}$)NiAsO samples with x = 0.1 and 0.2, measured in the zero-field-cooled
       (ZFC) and field-cooed (FC) processes. The superconducting fraction estimated at 2 K is beyond 40 \%. (b) The temperature dependence of resistivity of the sample
       with x = 0.1 under different magnetic fields. It is clear that the superconducting transition is broadened by using a magnetic field. The upper critical field is
       determined with the criterion $\rho=95\%\rho_n$ and shown as an inset of Fig.2(b). The solid line in the inset shows the theoretical fitting based on the GL theory (see text).}
       \label{figure3}
\end{figure}

\begin{figure}

       \includegraphics[width=8cm]{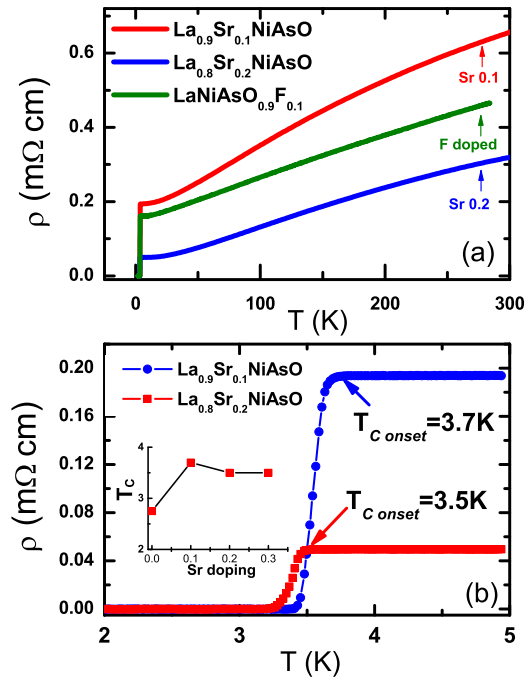}
       \caption{(Color online) (a)Temperature dependence of  resistivity in wide temperature region for samples with x=0.1 and 0.2, and the electron doped sample LaNiAs(O$_{0.9}$F$_{0.1}$).
       The normal state does not exhibit an anomaly which appears in the Fe-based system. (b) An enlarged view for the resistive transitions of the samples
       with x=0.1 and 0.2. The inset in Fig.3(b) presents the hole doping dependence of the transition temperature. Combining the data from the electron doped
       side, it is found that the curve of $T_c$ vs. hole and electron
       concentrations exhibits a symmetric behavior.}
       \label{figure4}
\end{figure}
 The DC magnetization data of (La$_{1-x}$Sr$_{x}$)NiAsO were shown
 in Fig.3(a). Fig.3(b) shows the temperature dependence of resistivity under different magnetic fields. A sharp transition with the width of about 0.4 K is observed at 3.7 K.
  By applying a magnetic field, the resistive transition curve broadens quickly showing a strong vortex flow behavior. But the onset transition point, which is close to the upper critical field, moves slowly with the magnetic
  field. This is similar to that observed in F-doped
  $LaFeAsO$\cite{ZhuXY}. Compared with the pure phase LaNiAsO with T$_c\approx$ 2.4 K\cite{hosono LaONiAs},  T$_c$
 of strontium substituted samples are improved to 3.7 K and 3.5 K for doping x=0.1 and 0.2,
 respectively. According to
   Ginzburg-Landau theory, zero temperature upper critical field
   H$_{c2}(0)$
   could be derived from the formula:
   H$_{c2}(T)$=H$_{c2}(0)$(1-t$^{2}$)/(1+t$^{2}$), where t is
   the normalized temperature T/T$_{c}$. It is found that the theoretical curve can fit the experimental data very well. The derived H$_{c2}(0)$ is
   found to be
   about 5.5 T, being close to that in the F-doped Ni-based system\cite{8}.

   Fig.4(a) shows the resistivity of (La$_{1-x}$Sr$_{x}$)NiAsO
   with x = 0.1, 0.2 from 2 K to 300 K at zero field. The resistivity in the normal state for all doping levels show metallic
   behavior. Near 3.7 K the resistivity of (La$_{0.9}$Sr$_{0.1}$)NiAsO
   drops sharply to zero, whereas, the
   resistivity of
   (La$_{0.8}$Sr$_{0.2}$)NiAsO drops at about 3.5 K with a similar transition width. For a better comparison, the resistivity of
   LaNiAs(O$_{0.9}$F$_{0.1}$) with T$_{c} \approx $3.8K was also shown in Fig.4(a). It is interesting to note that, at all doping levels the
normal state resistivity
   of the present Ni-based system
   exhibit no anomaly as found in the F-doped Fe-based system at
   about 150 K\cite{SDW}. A possible explanation is that there is a
   big difference in spin moment between Fe ion and Ni ion\cite{Ishibashi}.

In Fig.4(b) we show an enlarged view for the resistive transitions
for samples with x=0.1 and 0.2. The transition temperature of sample
x = 0.2 is about 3.5 K, which is very close to that of sample x=0.1,
but obviously higher than that of the undoped parent phase $LaNiAsO$
($T_c \approx 2.4 K$). Interestingly, if we plot the $T_c$ versus
the hole concentration, the curve exhibits a symmetric behavior with
the electron doped side\cite{8}. This behavior has also been found
in our original work for hole doped  $(La_{1-x}Sr_x)FeAsO$ system.
The similar behavior in both systems may suggest that the density of
states in the two sides of the Fermi energy is roughly symmetric.
Band structure calculation based on generalized gradient
approximation(GGA) functional has revealed that density of state is
roughly particle-hole symmetric in antiferromagnetic
LaFeAsO\cite{pickett}. As to LaNiAsO, also, band calculation using
GGA gave a similar result\cite{zhongfang}.

\begin{figure}

       \includegraphics[width=8cm]{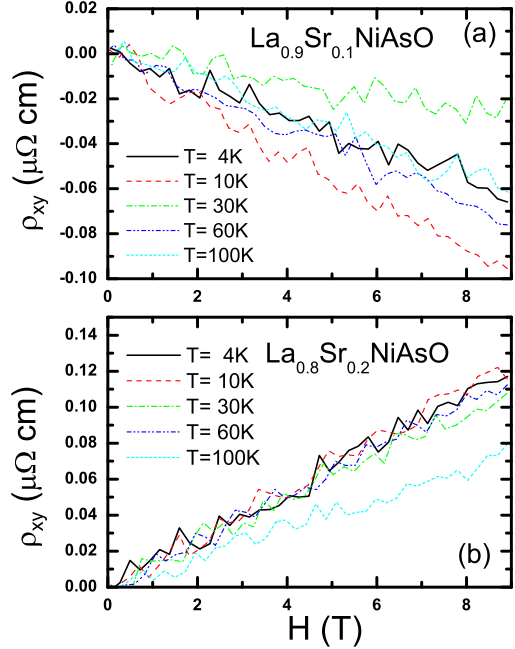}
       \caption{(Color online) Hall resistivity as a function of applied magnetic field for samples (La$_{1-x}$Sr$_{x}$)NiAsO, x = 0.1 (a) and 0.2 (b),
       respectively. The Hall resistivity is small in magnitude
       compared with the electron doped or undoped samples, indicating the gradual emergence of a hole-like conduction band.}
       \label{figure5}
\end{figure}

Since part of La$^{3+}$ are substituted by Sr$^{2+}$, hole typed
carriers are expected in our present Sr-doped system. A prove to
that by Hall effect measurements is necessary. Fig.5(a) and Fig.5(b)
show the Hall resistivity $\rho_{xy}$ for sample x = 0.1 and 0.2,
respectively. Interestingly, the sign of  $\rho_{xy}$ for x = 0.1 is
still negative, but quite close to zero. This is reasonable since
the parent phase $LaNiAsO$ is actually dominated by an electron-like
band\cite{8}, the Hall coefficient defined as $R_H=\rho_{xy}/H$ is
-5$\times10^{-10}m^{3}/C$ at 100 K for the undoped sample. This
means that holes are really introduced into the system by doping Sr.
By doping more Sr into the system, the Hall resistivity $\rho_{xy}$
becomes positive and hole-like charge carriers finally dominate the
conduction at the doping level x = 0.2. Fig.6 presents the Hall
coefficient  for two samples below 100 K. It is clear that
(La$_{0.9}$Sr$_{0.1}$)NiAsO has more electron-like charge carriers,
but the sample (La$_{0.8}$Sr$_{0.2}$)NiAsO shows clearly the
dominant conduction by hole-like charge carriers. Our data suggest
that with the substitution of La$^{3+}$ by Sr$^{2+}$, the conduction
by the electron-like band which appears for the undoped phase will
be prevailed over by the hole-like band, and superconductivity at
about 3.5-3.8 K occurs when the hole-like band dominates the
conduction. Moreover, such a phenomenon reflects that the Fermi
surface of $LaNiAsO$ does comprise of electron pockets and hole
pockets, thus the sign of charge carriers could be changed once the
contribution of hole pockets overwhelms that of electron pockets.

\begin{figure}

       \includegraphics[width=9cm]{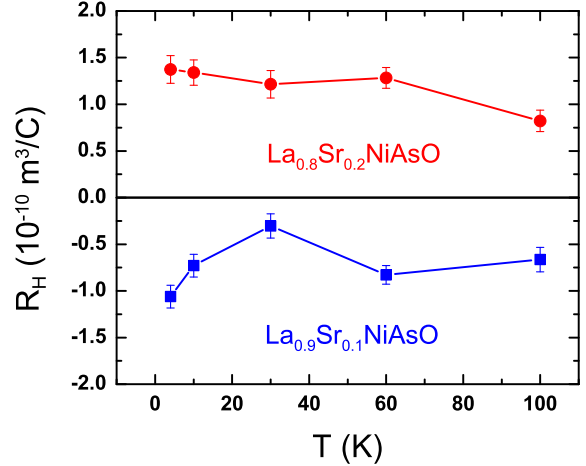}
       \caption{(Color online) Hall coefficients for samples (La$_{1-x}$Sr$_{x}$)NiAsO with x = 0.1 and 0.2. A sign change is obvious
       with increasing Sr content from 0.1 to 0.2 indicating a dominant conduction by hole-like charge carriers at x = 0.2.}
       \label{figure6}
\end{figure}

\begin{figure}

       \includegraphics[width=9cm]{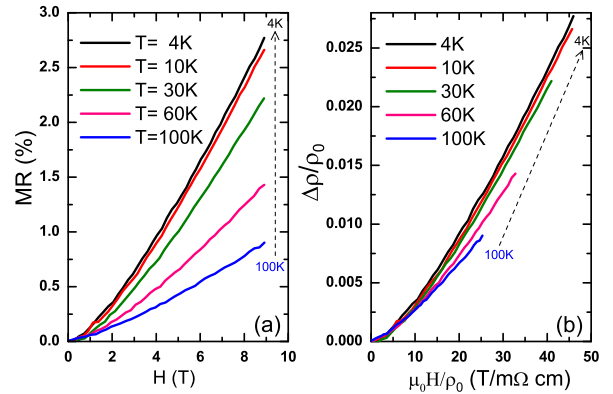}
       \caption{(Color online) (a) Field dependence of megnetoresistance $\triangle$$\rho$/$\rho$$_{0}$ at different temperatures for sample (La$_{0.9}$Sr$_{0.1}$)NiAsO .
       (b)Kohler plot at different temperatures,
       and obviously the Kohler rule is violated. }
       \label{figure7}
\end{figure}
Changing sign of Hall coefficient reveals the possible multiband
effect in the normal state of $(La_{1-x}Sr_x)NiAsO$, thus
magnetoresistance (MR) measurements and the suitability of Kohler's
rule is worth investigating. Fig.7(a) shows the magnetoresistance
versus the magnetic field at different temperatures. It is found
that the MR is about 2.7\% at 4K and 9T. Fig.7(b) shows the scaling
to the Kohler's rule, obviously the Kohler rule is violated. It is
believed that Kohler's law is conserved on single band metal with
symmetric Fermi surface topology. Therefore the magnetoresistance
effect and violation of Kohler's rule reveal that multiple bands
cross the Fermi surface. However taking account of the
polycrystalline samples that our experiments based on, we could not
exclude the skew scattering process caused by minor magnetic
impurities, thus single crystal samples of LaONiAs is strongly
desired.

In summary, by substituting La with Sr in $LaNiAsO$, a systematic
change of both the superconducting transition temperature and normal
state Hall coefficient are observed. First the transition
temperature is increased from 2.4 K to about 3.5 - 3.8 K with Sr
doping, meanwhile the Hall coefficient changes from negative to
positive. The curve of $T_c$ vs. the hole concentration exhibits a
symmetric behavior as the electron doped side, which may suggest a
roughly symmetric distribution of DOS above and below the Fermi
energy. Our data further support the conclusion that
superconductivity can be induced by hole doping.

\begin{acknowledgments}
We acknowledge the fruitful discussions with Yupeng Wang, Zidan Wang
and Tao Xiang. This work was financially supported by the NSF of
China, the MOST of China (973 Projects No. 2006CB601000, No.
2006CB921802 and 2006CB921300), and CAS (Project ITSNEM).
\end{acknowledgments}

\end{document}